\pdfoutput=1
\documentclass[11pt,twoside,a4paper,cmspaper,final,collab]{cms-tdr}
\def\svnVersion{27430M}\def\cmsCernNo{2010-052}\def\cmsCernDate{\today}\def\cmsMessage{Submitted to Physical Review Letters}

\begin{document}\cmsNoteHeader{EXO-10-005}
\hyphenation{env-iron-men-tal}
\hyphenation{had-ron-i-za-tion}
\hyphenation{cal-or-i-me-ter}
\hyphenation{de-vices}
\RCS$Revision: 27120 $
\RCS$HeadURL: svn+ssh://alverson@svn.cern.ch/reps/tdr2/papers/EXO-10-005/trunk/EXO-10-005.tex $
\RCS$Id: EXO-10-005.tex 27120 2010-12-16 13:36:52Z prumerio $
%
%
%

\providecommand {\etal}{\mbox{et al.}\xspace} 
\providecommand {\ie}{\mbox{i.e.}\xspace}     
\providecommand {\eg}{\mbox{e.g.}\xspace}     
\providecommand {\etc}{\mbox{etc.}\xspace}     
\providecommand {\vs}{\mbox{\sl vs.}\xspace}      
\providecommand {\mdash}{\ensuremath{\mathrm{-}}} 

\providecommand {\Lone}{Level-1\xspace} 
\providecommand {\Ltwo}{Level-2\xspace}
\providecommand {\Lthree}{Level-3\xspace}

\providecommand{\ACERMC} {\textsc{AcerMC}\xspace}
\providecommand{\ALPGEN} {{\textsc{alpgen}}\xspace}
\providecommand{\CHARYBDIS} {{\textsc{charybdis}}\xspace}
\providecommand{\CMKIN} {\textsc{cmkin}\xspace}
\providecommand{\CMSIM} {{\textsc{cmsim}}\xspace}
\providecommand{\CMSSW} {{\textsc{cmssw}}\xspace}
\providecommand{\COBRA} {{\textsc{cobra}}\xspace}
\providecommand{\COCOA} {{\textsc{cocoa}}\xspace}
\providecommand{\COMPHEP} {\textsc{CompHEP}\xspace}
\providecommand{\EVTGEN} {{\textsc{evtgen}}\xspace}
\providecommand{\FAMOS} {{\textsc{famos}}\xspace}
\providecommand{\GARCON} {\textsc{garcon}\xspace}
\providecommand{\GARFIELD} {{\textsc{garfield}}\xspace}
\providecommand{\GEANE} {{\textsc{geane}}\xspace}
\providecommand{\GEANTfour} {{\textsc{geant4}}\xspace}
\providecommand{\GEANTthree} {{\textsc{geant3}}\xspace}
\providecommand{\GEANT} {{\textsc{geant}}\xspace}
\providecommand{\HDECAY} {\textsc{hdecay}\xspace}
\providecommand{\HERWIG} {{\textsc{herwig}}\xspace}
\providecommand{\HIGLU} {{\textsc{higlu}}\xspace}
\providecommand{\HIJING} {{\textsc{hijing}}\xspace}
\providecommand{\IGUANA} {\textsc{iguana}\xspace}
\providecommand{\ISAJET} {{\textsc{isajet}}\xspace}
\providecommand{\ISAPYTHIA} {{\textsc{isapythia}}\xspace}
\providecommand{\ISASUGRA} {{\textsc{isasugra}}\xspace}
\providecommand{\ISASUSY} {{\textsc{isasusy}}\xspace}
\providecommand{\ISAWIG} {{\textsc{isawig}}\xspace}
\providecommand{\MADGRAPH} {\textsc{MadGraph}\xspace}
\providecommand{\MCATNLO} {\textsc{mc@nlo}\xspace}
\providecommand{\MCFM} {\textsc{mcfm}\xspace}
\providecommand{\MILLEPEDE} {{\textsc{millepede}}\xspace}
\providecommand{\ORCA} {{\textsc{orca}}\xspace}
\providecommand{\OSCAR} {{\textsc{oscar}}\xspace}
\providecommand{\PHOTOS} {\textsc{photos}\xspace}
\providecommand{\PROSPINO} {\textsc{prospino}\xspace}
\providecommand{\PYTHIA} {{\textsc{pythia}}\xspace}
\providecommand{\SHERPA} {{\textsc{sherpa}}\xspace}
\providecommand{\TAUOLA} {\textsc{tauola}\xspace}
\providecommand{\TOPREX} {\textsc{TopReX}\xspace}
\providecommand{\XDAQ} {{\textsc{xdaq}}\xspace}

\providecommand {\DZERO}{D\O\xspace}     


\providecommand{\de}{\ensuremath{^\circ}}
\providecommand{\ten}[1]{\ensuremath{\times \text{10}^\text{#1}}}
\providecommand{\unit}[1]{\ensuremath{\text{\,#1}}\xspace}
\providecommand{\mum}{\ensuremath{\,\mu\text{m}}\xspace}
\providecommand{\micron}{\ensuremath{\,\mu\text{m}}\xspace}
\providecommand{\cm}{\ensuremath{\,\text{cm}}\xspace}
\providecommand{\mm}{\ensuremath{\,\text{mm}}\xspace}
\providecommand{\mus}{\ensuremath{\,\mu\text{s}}\xspace}
\providecommand{\keV}{\ensuremath{\,\text{ke\hspace{-.08em}V}}\xspace}
\providecommand{\MeV}{\ensuremath{\,\text{Me\hspace{-.08em}V}}\xspace}
\providecommand{\GeV}{\ensuremath{\,\text{Ge\hspace{-.08em}V}}\xspace}
\providecommand{\gev}{\GeV}
\providecommand{\TeV}{\ensuremath{\,\text{Te\hspace{-.08em}V}}\xspace}
\providecommand{\PeV}{\ensuremath{\,\text{Pe\hspace{-.08em}V}}\xspace}
\providecommand{\keVc}{\ensuremath{{\,\text{ke\hspace{-.08em}V\hspace{-0.16em}/\hspace{-0.08em}}c}}\xspace}
\providecommand{\MeVc}{\ensuremath{{\,\text{Me\hspace{-.08em}V\hspace{-0.16em}/\hspace{-0.08em}}c}}\xspace}
\providecommand{\GeVc}{\ensuremath{{\,\text{Ge\hspace{-.08em}V\hspace{-0.16em}/\hspace{-0.08em}}c}}\xspace}
\providecommand{\TeVc}{\ensuremath{{\,\text{Te\hspace{-.08em}V\hspace{-0.16em}/\hspace{-0.08em}}c}}\xspace}
\providecommand{\keVcc}{\ensuremath{{\,\text{ke\hspace{-.08em}V\hspace{-0.16em}/\hspace{-0.08em}}c^\text{2}}}\xspace}
\providecommand{\MeVcc}{\ensuremath{{\,\text{Me\hspace{-.08em}V\hspace{-0.16em}/\hspace{-0.08em}}c^\text{2}}}\xspace}
\providecommand{\GeVcc}{\ensuremath{{\,\text{Ge\hspace{-.08em}V\hspace{-0.16em}/\hspace{-0.08em}}c^\text{2}}}\xspace}
\providecommand{\TeVcc}{\ensuremath{{\,\text{Te\hspace{-.08em}V\hspace{-0.16em}/\hspace{-0.08em}}c^\text{2}}}\xspace}

\providecommand{\pbinv} {\mbox{\ensuremath{\,\text{pb}^\text{$-$1}}}\xspace}
\providecommand{\fbinv} {\mbox{\ensuremath{\,\text{fb}^\text{$-$1}}}\xspace}
\providecommand{\nbinv} {\mbox{\ensuremath{\,\text{nb}^\text{$-$1}}}\xspace}
\providecommand{\percms}{\ensuremath{\,\text{cm}^\text{$-$2}\,\text{s}^\text{$-$1}}\xspace}
\providecommand{\lumi}{\ensuremath{\mathcal{L}}\xspace}
\providecommand{\Lumi}{\ensuremath{\mathcal{L}}\xspace}
%
\providecommand{\LvLow}  {\ensuremath{\mathcal{L}=\text{10}^\text{32}\,\text{cm}^\text{$-$2}\,\text{s}^\text{$-$1}}\xspace}
\providecommand{\LLow}   {\ensuremath{\mathcal{L}=\text{10}^\text{33}\,\text{cm}^\text{$-$2}\,\text{s}^\text{$-$1}}\xspace}
\providecommand{\lowlumi}{\ensuremath{\mathcal{L}=\text{2}\times \text{10}^\text{33}\,\text{cm}^\text{$-$2}\,\text{s}^\text{$-$1}}\xspace}
\providecommand{\LMed}   {\ensuremath{\mathcal{L}=\text{2}\times \text{10}^\text{33}\,\text{cm}^\text{$-$2}\,\text{s}^\text{$-$1}}\xspace}
\providecommand{\LHigh}  {\ensuremath{\mathcal{L}=\text{10}^\text{34}\,\text{cm}^\text{$-$2}\,\text{s}^\text{$-$1}}\xspace}
\providecommand{\hilumi} {\ensuremath{\mathcal{L}=\text{10}^\text{34}\,\text{cm}^\text{$-$2}\,\text{s}^\text{$-$1}}\xspace}


\providecommand{\PT}{\ensuremath{p_{\mathrm{T}}}\xspace}
\providecommand{\pt}{\ensuremath{p_{\mathrm{T}}}\xspace}
\providecommand{\ET}{\ensuremath{E_{\mathrm{T}}}\xspace}
\providecommand{\HT}{\ensuremath{H_{\mathrm{T}}}\xspace}
\providecommand{\et}{\ensuremath{E_{\mathrm{T}}}\xspace}
\providecommand{\Em}{\ensuremath{E\hspace{-0.6em}/}\xspace}
\providecommand{\Pm}{\ensuremath{p\hspace{-0.5em}/}\xspace}
\providecommand{\PTm}{\ensuremath{{p}_\mathrm{T}\hspace{-1.02em}/}\xspace}
\providecommand{\PTslash}{\ensuremath{{p}_\mathrm{T}\hspace{-1.02em}/}\xspace}
\providecommand{\ETm}{\ensuremath{E_{\mathrm{T}}^{\text{miss}}}\xspace}
\providecommand{\ETslash}{\ensuremath{E_{\mathrm{T}}\hspace{-1.1em}/}\xspace}
\providecommand{\MET}{\ensuremath{E_{\mathrm{T}}^{\text{miss}}}\xspace}
\providecommand{\ETmiss}{\ensuremath{E_{\mathrm{T}}^{\text{miss}}}\xspace}
\providecommand{\VEtmiss}{\ensuremath{{\vec E}_{\mathrm{T}}^{\text{miss}}}\xspace}

\providecommand{\dd}[2]{\ensuremath{\frac{\mathrm{d} #1}{\mathrm{d} #2}}}

\ifthenelse{\boolean{cms@italic}}{\newcommand{\cmsSymbolFace}{\relax}}{\newcommand{\cmsSymbolFace}{\mathrm}}

\providecommand{\zp}{\ensuremath{\cmsSymbolFace{Z}^\prime}\xspace}
\providecommand{\JPsi}{\ensuremath{\cmsSymbolFace{J}\hspace{-.08em}/\hspace{-.14em}\psi}\xspace}
\providecommand{\Z}{\ensuremath{\cmsSymbolFace{Z}}\xspace}
\providecommand{\ttbar}{\ensuremath{\cmsSymbolFace{t}\overline{\cmsSymbolFace{t}}}\xspace}


\providecommand{\AFB}{\ensuremath{A_\text{FB}}\xspace}
\providecommand{\wangle}{\ensuremath{\sin^{2}\theta_{\text{eff}}^\text{lept}(M^2_\Z)}\xspace}
\providecommand{\stat}{\ensuremath{\,\text{(stat.)}}\xspace}
\providecommand{\syst}{\ensuremath{\,\text{(syst.)}}\xspace}
\providecommand{\kt}{\ensuremath{k_{\mathrm{T}}}\xspace}

\providecommand{\BC}{\ensuremath{\mathrm{B_{c}}}\xspace}
\providecommand{\bbarc}{\ensuremath{\mathrm{\overline{b}c}}\xspace}
\providecommand{\bbbar}{\ensuremath{\mathrm{b\overline{b}}}\xspace}
\providecommand{\ccbar}{\ensuremath{\mathrm{c\overline{c}}}\xspace}
\providecommand{\bspsiphi}{\ensuremath{\mathrm{B_s} \to \JPsi\, \phi}\xspace}
\providecommand{\EE}{\ensuremath{\mathrm{e^+e^-}}\xspace}
\providecommand{\MM}{\ensuremath{\mu^+\mu^-}\xspace}
\providecommand{\TT}{\ensuremath{\tau^+\tau^-}\xspace}

\providecommand{\HGG}{\ensuremath{\mathrm{H}\to\gamma\gamma}}
\providecommand{\GAMJET}{\ensuremath{\gamma + \text{jet}}}
\providecommand{\PPTOJETS}{\ensuremath{\mathrm{pp}\to\text{jets}}}
\providecommand{\PPTOGG}{\ensuremath{\mathrm{pp}\to\gamma\gamma}}
\providecommand{\PPTOGAMJET}{\ensuremath{\mathrm{pp}\to\gamma + \mathrm{jet}}}
\providecommand{\MH}{\ensuremath{M_{\mathrm{H}}}}
\providecommand{\RNINE}{\ensuremath{R_\mathrm{9}}}
\providecommand{\DR}{\ensuremath{\Delta R}}

%

\providecommand{\ga}{\ensuremath{\gtrsim}}
\providecommand{\la}{\ensuremath{\lesssim}}
\providecommand{\swsq}{\ensuremath{\sin^2\theta_\mathrm{W}}\xspace}
\providecommand{\cwsq}{\ensuremath{\cos^2\theta_\mathrm{W}}\xspace}
\providecommand{\tanb}{\ensuremath{\tan\beta}\xspace}
\providecommand{\tanbsq}{\ensuremath{\tan^{2}\beta}\xspace}
\providecommand{\sidb}{\ensuremath{\sin 2\beta}\xspace}
\providecommand{\alpS}{\ensuremath{\alpha_S}\xspace}
\providecommand{\alpt}{\ensuremath{\tilde{\alpha}}\xspace}

\providecommand{\QL}{\ensuremath{\mathrm{Q}_\mathrm{L}}\xspace}
\providecommand{\sQ}{\ensuremath{\tilde{\mathrm{Q}}}\xspace}
\providecommand{\sQL}{\ensuremath{\tilde{\mathrm{Q}}_\mathrm{L}}\xspace}
\providecommand{\ULC}{\ensuremath{\mathrm{U}_\mathrm{L}^\mathrm{C}}\xspace}
\providecommand{\sUC}{\ensuremath{\tilde{\mathrm{U}}^\mathrm{C}}\xspace}
\providecommand{\sULC}{\ensuremath{\tilde{\mathrm{U}}_\mathrm{L}^\mathrm{C}}\xspace}
\providecommand{\DLC}{\ensuremath{\mathrm{D}_\mathrm{L}^\mathrm{C}}\xspace}
\providecommand{\sDC}{\ensuremath{\tilde{\mathrm{D}}^\mathrm{C}}\xspace}
\providecommand{\sDLC}{\ensuremath{\tilde{\mathrm{D}}_\mathrm{L}^\mathrm{C}}\xspace}
\providecommand{\LL}{\ensuremath{\mathrm{L}_\mathrm{L}}\xspace}
\providecommand{\sL}{\ensuremath{\tilde{\mathrm{L}}}\xspace}
\providecommand{\sLL}{\ensuremath{\tilde{\mathrm{L}}_\mathrm{L}}\xspace}
\providecommand{\ELC}{\ensuremath{\mathrm{E}_\mathrm{L}^\mathrm{C}}\xspace}
\providecommand{\sEC}{\ensuremath{\tilde{\mathrm{E}}^\mathrm{C}}\xspace}
\providecommand{\sELC}{\ensuremath{\tilde{\mathrm{E}}_\mathrm{L}^\mathrm{C}}\xspace}
\providecommand{\sEL}{\ensuremath{\tilde{\mathrm{E}}_\mathrm{L}}\xspace}
\providecommand{\sER}{\ensuremath{\tilde{\mathrm{E}}_\mathrm{R}}\xspace}
\providecommand{\sFer}{\ensuremath{\tilde{\mathrm{f}}}\xspace}
\providecommand{\sQua}{\ensuremath{\tilde{\mathrm{q}}}\xspace}
\providecommand{\sUp}{\ensuremath{\tilde{\mathrm{u}}}\xspace}
\providecommand{\suL}{\ensuremath{\tilde{\mathrm{u}}_\mathrm{L}}\xspace}
\providecommand{\suR}{\ensuremath{\tilde{\mathrm{u}}_\mathrm{R}}\xspace}
\providecommand{\sDw}{\ensuremath{\tilde{\mathrm{d}}}\xspace}
\providecommand{\sdL}{\ensuremath{\tilde{\mathrm{d}}_\mathrm{L}}\xspace}
\providecommand{\sdR}{\ensuremath{\tilde{\mathrm{d}}_\mathrm{R}}\xspace}
\providecommand{\sTop}{\ensuremath{\tilde{\mathrm{t}}}\xspace}
\providecommand{\stL}{\ensuremath{\tilde{\mathrm{t}}_\mathrm{L}}\xspace}
\providecommand{\stR}{\ensuremath{\tilde{\mathrm{t}}_\mathrm{R}}\xspace}
\providecommand{\stone}{\ensuremath{\tilde{\mathrm{t}}_1}\xspace}
\providecommand{\sttwo}{\ensuremath{\tilde{\mathrm{t}}_2}\xspace}
\providecommand{\sBot}{\ensuremath{\tilde{\mathrm{b}}}\xspace}
\providecommand{\sbL}{\ensuremath{\tilde{\mathrm{b}}_\mathrm{L}}\xspace}
\providecommand{\sbR}{\ensuremath{\tilde{\mathrm{b}}_\mathrm{R}}\xspace}
\providecommand{\sbone}{\ensuremath{\tilde{\mathrm{b}}_1}\xspace}
\providecommand{\sbtwo}{\ensuremath{\tilde{\mathrm{b}}_2}\xspace}
\providecommand{\sLep}{\ensuremath{\tilde{\mathrm{l}}}\xspace}
\providecommand{\sLepC}{\ensuremath{\tilde{\mathrm{l}}^\mathrm{C}}\xspace}
\providecommand{\sEl}{\ensuremath{\tilde{\mathrm{e}}}\xspace}
\providecommand{\sElC}{\ensuremath{\tilde{\mathrm{e}}^\mathrm{C}}\xspace}
\providecommand{\seL}{\ensuremath{\tilde{\mathrm{e}}_\mathrm{L}}\xspace}
\providecommand{\seR}{\ensuremath{\tilde{\mathrm{e}}_\mathrm{R}}\xspace}
\providecommand{\snL}{\ensuremath{\tilde{\nu}_L}\xspace}
\providecommand{\sMu}{\ensuremath{\tilde{\mu}}\xspace}
\providecommand{\sNu}{\ensuremath{\tilde{\nu}}\xspace}
\providecommand{\sTau}{\ensuremath{\tilde{\tau}}\xspace}
\providecommand{\Glu}{\ensuremath{\mathrm{g}}\xspace}
\providecommand{\sGlu}{\ensuremath{\tilde{\mathrm{g}}}\xspace}
\providecommand{\Wpm}{\ensuremath{\mathrm{W}^{\pm}}\xspace}
\providecommand{\sWpm}{\ensuremath{\tilde{\mathrm{W}}^{\pm}}\xspace}
\providecommand{\Wz}{\ensuremath{\mathrm{W}^{0}}\xspace}
\providecommand{\sWz}{\ensuremath{\tilde{\mathrm{W}}^{0}}\xspace}
\providecommand{\sWino}{\ensuremath{\tilde{\mathrm{W}}}\xspace}
\providecommand{\Bz}{\ensuremath{\mathrm{B}^{0}}\xspace}
\providecommand{\sBz}{\ensuremath{\tilde{\mathrm{B}}^{0}}\xspace}
\providecommand{\sBino}{\ensuremath{\tilde{\mathrm{B}}}\xspace}
\providecommand{\Zz}{\ensuremath{\mathrm{Z}^{0}}\xspace}
\providecommand{\sZino}{\ensuremath{\tilde{\mathrm{Z}}^{0}}\xspace}
\providecommand{\sGam}{\ensuremath{\tilde{\gamma}}\xspace}
\providecommand{\chiz}{\ensuremath{\tilde{\chi}^{0}}\xspace}
\providecommand{\chip}{\ensuremath{\tilde{\chi}^{+}}\xspace}
\providecommand{\chim}{\ensuremath{\tilde{\chi}^{-}}\xspace}
\providecommand{\chipm}{\ensuremath{\tilde{\chi}^{\pm}}\xspace}
\providecommand{\Hone}{\ensuremath{\mathrm{H}_\mathrm{d}}\xspace}
\providecommand{\sHone}{\ensuremath{\tilde{\mathrm{H}}_\mathrm{d}}\xspace}
\providecommand{\Htwo}{\ensuremath{\mathrm{H}_\mathrm{u}}\xspace}
\providecommand{\sHtwo}{\ensuremath{\tilde{\mathrm{H}}_\mathrm{u}}\xspace}
\providecommand{\sHig}{\ensuremath{\tilde{\mathrm{H}}}\xspace}
\providecommand{\sHa}{\ensuremath{\tilde{\mathrm{H}}_\mathrm{a}}\xspace}
\providecommand{\sHb}{\ensuremath{\tilde{\mathrm{H}}_\mathrm{b}}\xspace}
\providecommand{\sHpm}{\ensuremath{\tilde{\mathrm{H}}^{\pm}}\xspace}
\providecommand{\hz}{\ensuremath{\mathrm{h}^{0}}\xspace}
\providecommand{\Hz}{\ensuremath{\mathrm{H}^{0}}\xspace}
\providecommand{\Az}{\ensuremath{\mathrm{A}^{0}}\xspace}
\providecommand{\Hpm}{\ensuremath{\mathrm{H}^{\pm}}\xspace}
\providecommand{\sGra}{\ensuremath{\tilde{\mathrm{G}}}\xspace}
\providecommand{\mtil}{\ensuremath{\tilde{m}}\xspace}
\providecommand{\rpv}{\ensuremath{\rlap{\kern.2em/}R}\xspace}
\providecommand{\LLE}{\ensuremath{LL\bar{E}}\xspace}
\providecommand{\LQD}{\ensuremath{LQ\bar{D}}\xspace}
\providecommand{\UDD}{\ensuremath{\overline{UDD}}\xspace}
\providecommand{\Lam}{\ensuremath{\lambda}\xspace}
\providecommand{\Lamp}{\ensuremath{\lambda'}\xspace}
\providecommand{\Lampp}{\ensuremath{\lambda''}\xspace}
\providecommand{\spinbd}[2]{\ensuremath{\bar{#1}_{\dot{#2}}}\xspace}

\providecommand{\MD}{\ensuremath{{M_\mathrm{D}}}\xspace}
\providecommand{\Mpl}{\ensuremath{{M_\mathrm{Pl}}}\xspace}
\providecommand{\Rinv} {\ensuremath{{R}^{-1}}\xspace} 
\def\zee{\ensuremath{\Z \to \cmsSymbolFace{e}\cmsSymbolFace{e}}\xspace}
\def\zjets{\ensuremath{\Z/\gamma}+jets\xspace}
\def\aeta{\ensuremath{|\eta|}\xspace}
\def\pgev{\GeV}
\def\mgev{\GeV}
\def\mee{\ensuremath{M_{\text{ee}}}\xspace}
\def\st{\ensuremath{S_{\text{T}}}\xspace}
\def\pt{\ensuremath{p_{\rm T}}\xspace}
\def \LQ {\mbox{LQ}\xspace}
\def \Lint {$\mathcal{L}_{int}$\xspace}
\def \GeVmass {GeV\xspace}
\def \GeVmom {GeV\xspace}
\def \MLQ {$M_{\rm LQ}$\xspace}
\def \pp {$\Pp \Pp$\xspace}
\def \sqrts {$\sqrt{s}=7$~TeV\xspace}
\def \etmiss {\big\slash\hspace{-1.6ex}{E_\text{T}}}
\def \lumi {33.2~pb$^{-1}$\xspace} 
\def \lumiwitherror {33.2$\pm$3.7~pb$^{-1}$\xspace} 
\def \lumiOpt {35~pb$^{-1}$\xspace} 
\def \LQmassLimitBetaOne {$384$~GeV\xspace} 
\def \LQmassLimitBetaOneWithTheoryError {$370$~GeV\xspace} 
\def \LQmassLimitBetaHalf {$XXX$~GeV\xspace} 
\def \NEventsWithTwoEles {XXX\xspace}
\def \MeeFinalCut {125\xspace}
\def \rescaleFactor {$XXX\pmXXX$\xspace} 
\def \NpreselDATA {$N_{\rm Data}=XXX$\xspace} 
\def \NpreselMC {$N_{\rm MC}=XXX\pmXXX$\xspace} 
\def \ZbkgPrelimStCut {250~GeV\xspace}  
\def \ContaminationAtZpeak {$3\%$\xspace} 
\def \ZbkgUncertainty {$12\%$\xspace}  
\def \ZbkgRatioAtZPeak {1.20$\pm$0.14\xspace}  
\def \ZbkgUncertaintyTwo {$XXX\%$\xspace}  
\def \QCDbkg {$<1\%$\xspace} 
\def \uncertaintyQCD {$20\%$\xspace} 
\def \Npileup {XXX\xspace} 
\def \ttbarRelErr {$41\%$\xspace} 
\def \NemujjData {XXX\xspace} 
\def \NemujjMC {$XXX\pmXXX$\xspace} 
\def \emujjContamination {$XXX\%$\xspace}  
\def \maxInstLumi {$2~\times~10^{32}~$cm$^{-2}$s$^{-1}$\xspace} 
\def \twoEleEffic {from 58.7\% to 68.0\%\xspace} 
\def \MLQforSystematics {$300$\xspace} 
\def \MLQforMejPlotInAppendix {$300$\xspace} 
\def \LQMassesAnalyzed{from 200 to 500 GeV\xspace}
\def \PDFUncertaintyOnLQCrossSection{from 8 to 22\% for LQ masses from 200 to 500 GeV\xspace}
\def \theoreticalUncertaintyOnLQCrossSection{14-15\% for all LQ masses considered\xspace}
\cmsNoteHeader{EXO-10-005} 
\title{Search for Pair Production of First-Generation Scalar Leptoquarks
in $\Pp \Pp$  Collisions at $\sqrt{s}= 7$~TeV} 
\address[Maryland]{U. Maryland}
\author[cern]{The CMS Collaboration}
\date{\today}
\abstract{
A search for pair production of first-generation scalar leptoquarks
is performed in the final state containing two electrons and two
jets using proton-proton collision data at $\sqrt{s}=7$~TeV.
The data sample used corresponds to an
integrated luminosity of 33~pb$^{-1}$ collected with the CMS
detector at the CERN LHC. The number of observed events is in good
agreement with the predictions for the standard model background
processes, and an upper limit is set on the leptoquark pair production cross
section times $\beta^2$ as a function of the leptoquark mass, where $\beta$ is
the branching fraction of the leptoquark decay to an electron and a quark.
A 95\% confidence level lower limit is set on the mass of a first-generation
scalar leptoquark at 384~GeV for $\beta=1$, which is the most stringent direct
limit to date.
}
\hypersetup{%
pdfauthor={CMS Collaboration},%
pdftitle={Search for Pair Production of First Generation Scalar Leptoquarks
in pp Collisions at sqrt s = 7 TeV},%
pdfsubject={CMS},%
pdfkeywords={CMS, physics}}

\maketitle 
Although the standard model (SM) of fundamental particles and their
interactions is in excellent agreement with most collider data,
there are compelling reasons to believe new physics should
appear at high energy scales.  Some well-motivated theories of
physics beyond the SM, including
grand unified theories \cite{Pati:1974yy},
composite models \cite{Schrempp:1984nj}, technicolor \cite{Dimopoulos:1979es,Dimopoulos:1979sp,Farhi:1980xs},
and super\-string-inspired $E_6$ models \cite{Hewett:1988xc},
postulate the existence of a symmetry, beyond that of the SM,
relating quarks and leptons and
implying the existence of new bosons, called leptoquarks (LQ).
An LQ carries colour, has fractional electric charge, can
have spin 0 (scalar) or spin 1 (vector),
and couples to a lepton and a quark with coupling strength
$\lambda$.
An LQ would
decay to a charged lepton and a quark, with an unknown branching fraction
$\beta$, or a neutrino and a quark, with branching fraction
$1-\beta$.
A review of LQ phenomenology and searches can be found in \cite{Acosta:1999ws,Hewett:1997ce}.
Constraints from experiments sensitive to flavour-changing
neutral currents, lepton-family-number violation,
and other rare processes \cite{Buchmuller:1986iq}
favour LQs that couple to quarks and leptons within the same SM generation,
for LQ masses accessible at current colliders.
The first-generation scalar LQs studied in this Letter couple only
to an electron or an electron neutrino and a light quark.
Measurements at electron-proton colliders constrain
the coupling $\lambda$ to be comparable to or less than
the electromagnetic coupling $\lambda_{\rm EM}\equiv \sqrt{4\pi\alpha_{\rm EM}} \approx 0.3$,
for a first generation LQ mass, $M_{\rm LQ}$, less than 300~\GeVmass~\cite{Aktas:2005pr,Chekanov:2003af}.
Prior to this work, the \DZERO Collaboration set the most stringent limit
for a broad range of the coupling $\lambda$ on the mass of the
first-generation scalar LQ, namely, $M_\text{LQ}>299$~GeV for $\beta=1$ \cite{Abazov:2009gf}.

This Letter presents the results of a search for pair production of
first-generation scalar LQs using events containing two electrons and two jets
from a data sample of \pp collisions at $\sqrt{s} = 7$~TeV collected in 2010 with the
Compact Muon Solenoid (CMS) detector at the LHC.
The data sample corresponds to an integrated luminosity of \lumiwitherror.
In \pp~collisions at this energy, LQs are predominantly produced in pairs
via gluon-gluon fusion and quark-antiquark annihilation with a
cross section that depends on the strong coupling
constant $\alpha_{\rm s}$ but is nearly independent of $\lambda$.
This cross section depends on the spin and the mass
of the LQ and has been calculated including Next-to-Leading-Order (NLO)
Quantum Chromodynamics (QCD) corrections~\cite{PhysRevD.71.057503}.
In this study we did not consider possible
contributions from single LQ production, which has a cross section
that is dependent on $\lambda$.

The CMS experiment~\cite{:2008zzk} 
uses a right-handed coordinate system, with the origin at the
nominal interaction point, the $x$-axis pointing to the center of
the LHC ring, the $y$-axis pointing up (perpendicular to the LHC plane),
and the $z$-axis along the anticlockwise-beam direction. The polar
angle, $\theta$, is measured from the positive $z$-axis and the
azimuthal angle, $\phi$, is measured in the $x$-$y$ plane.
The pseudorapidity is given by $\eta = -\ln(\tan\theta/2)$. 
The central feature of the CMS apparatus is a
superconducting solenoid, of 6~m internal diameter, providing a field
of 3.8~T. Within the field volume are the silicon pixel and strip
tracker, the crystal electromagnetic calorimeter (ECAL), 
which includes a silicon sensor preshower detector in front of the ECAL endcaps, 
and the brass/scintillator hadron calorimeter (HCAL). Muons are measured in
gas-ionization detectors embedded in the steel return yoke. In
addition to the barrel and endcap detectors, CMS has extensive forward
calorimetry.
The ECAL has an ultimate energy resolution of better than 0.5\% 
for unconverted photons with transverse energies above 100 GeV. 
The energy resolution is 3\% or better for the range of electron energies 
relevant for this analysis.
The HCAL, when combined with the ECAL, measures jets with a
resolution $\Delta E/E \approx 100\,\% / \sqrt{E\mathrm{[GeV]}} \oplus 5\,\%$. 
The inner tracker measures charged particles within $|\eta| < 2.5$ 
and provides an impact parameter resolution of $\sim$\,15~$\mu$m and a
transverse momentum ($p_{\rm T}$) resolution of about 1.5\,\% for
100~GeV particles.
The relative luminosity is measured using the forward calorimeters.
Collision events were selected by a first level trigger made of a system of
fast electronics and a higher level trigger that consists
of a farm of commercial CPUs running a version of the offline
reconstruction optimized for fast processing.

Events used in this analysis are collected with an efficiency of 100\% by single and double electron triggers
with various thresholds depending on the instantaneous luminosity.
Offline, events are required to contain at least one primary
vertex with $z$-position within 24 cm of the nominal center of the detector.
Electron candidates are required to have an
electromagnetic cluster in ECAL
that is spatially matched to a reconstructed track
in the central tracking system in both $\eta$ and $\phi$.  Electron
candidates are further required to have a shower shape consistent with
that of an electromagnetic shower,
have a ratio between the hadronic and electromagnetic energy of less than 5\%,
and be isolated within a cone $\sqrt{\Delta\eta^2+\Delta\phi^2}<0.3$
from other energy
deposits in the calorimeter and from additional reconstructed tracks
(beyond the matched track) in the central tracking system.
More information about electron triggering and identification at CMS
can be found elsewhere \cite{EGM-10-004}.
Jets, the experimental signature of the hadronization of partons,
are reconstructed in this analysis from calorimetry information
by the anti-k$_\text{T}$ algorithm \cite{antikt} with the distance parameter set to 0.5.
The energy response of the jets is adjusted by applying a correction determined
from Monte Carlo (MC) simulated events and a
residual correction derived from data by analyzing the \pt balance
in di-jet events \cite{PAS-JME-10-010}.

The collision data were compared to samples of MC generated
events, where the response of the detector was simulated using
\GEANTfour \cite{GEANT4}.
The selection procedure as well as the electron and jet reconstructions
described for the data are also applied to the MC simulation samples.
Signal samples for LQ masses \LQMassesAnalyzed
were generated with \PYTHIA \cite{pythia}, version 6.422, tune D6T \cite{Field:2008zz,Field:2010su}. 

\begin{figure}[htbp]
  \begin{center}
    \includegraphics[width=0.45\textwidth]{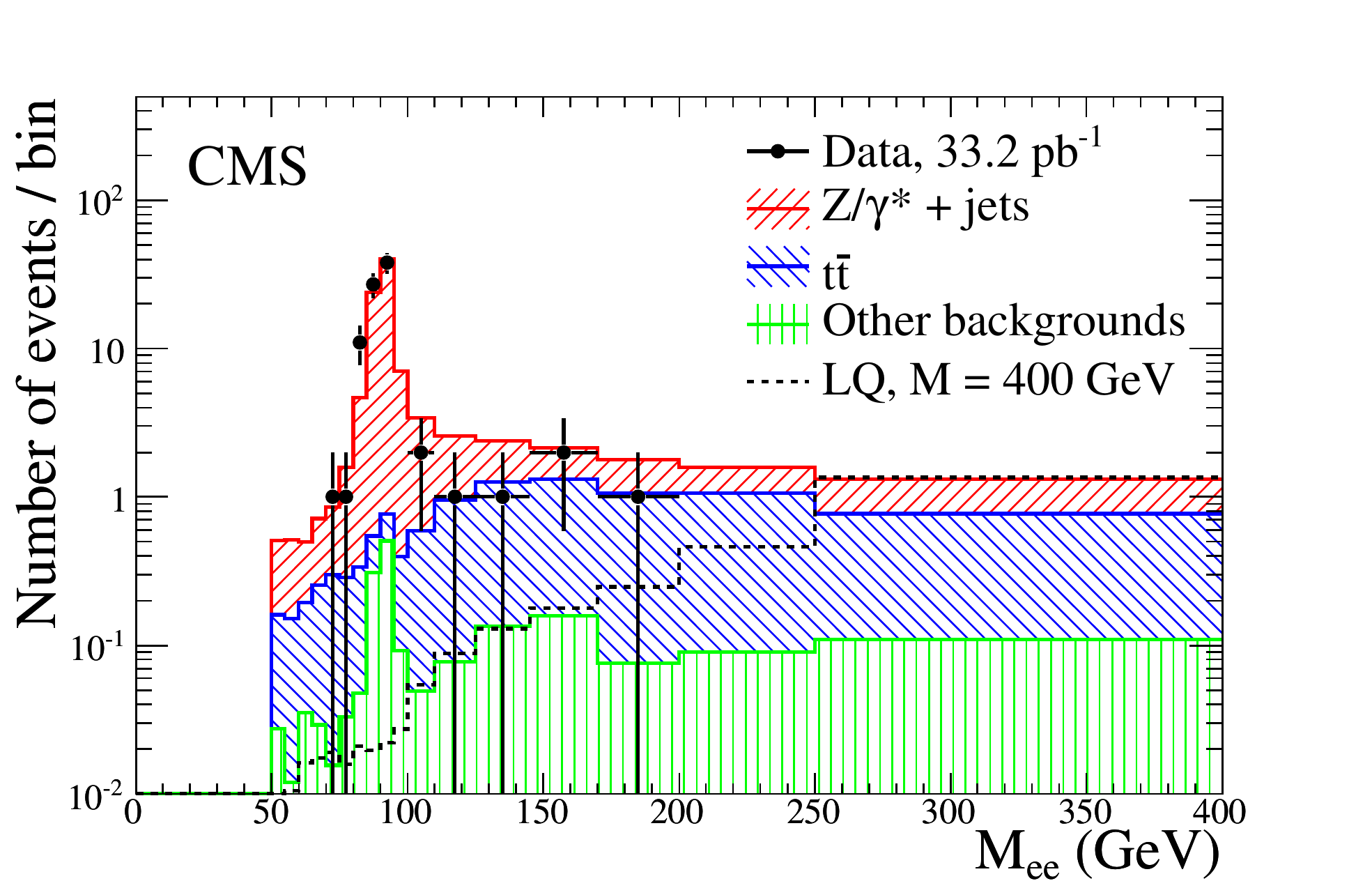}
    \includegraphics[width=0.45\textwidth]{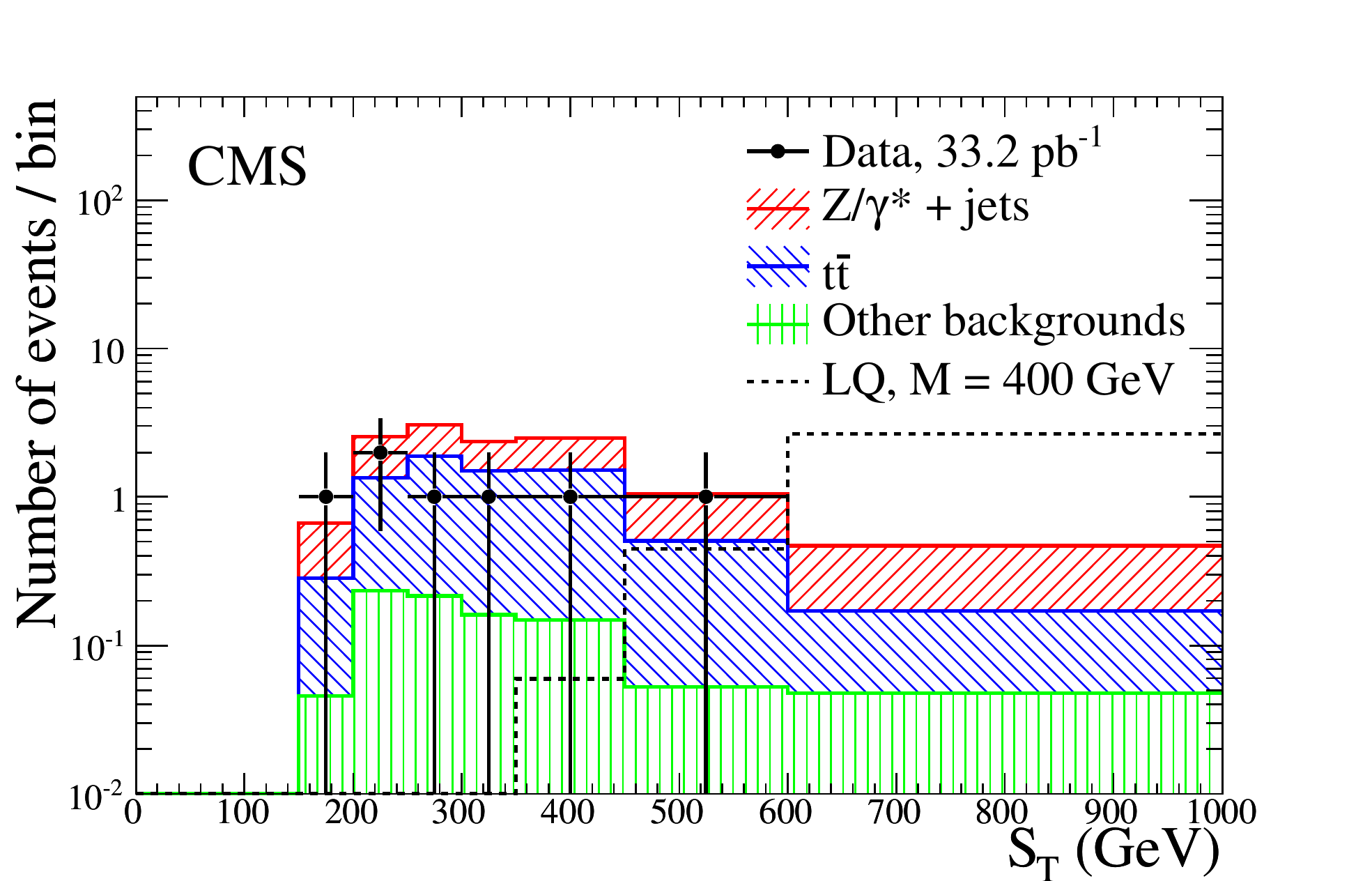}
    \caption{On the left: the \mee distribution 
	     for events that have passed the pre-selection requirements.
	     On the right: the \st distribution for events that have passed the pre-selection requirement,
	     except the pre-selection requirement on \st itself (\st$>250$~GeV),
	     and have \mee$>\MeeFinalCut$~GeV.
	     The MC distributions for the signal ($\beta=1$) 
	     and the contributing backgrounds are shown.
	     The \zjets MC has been normalized as described in the text.
	     Other backgrounds include $\PW$+jets, di-boson, and single top.
	     All background histograms are cumulative.}
    \label{fig:MeeAfterPresel}
  \end{center}
\end{figure}

An initial sample containing at least two electrons and at least two jets is selected.
The dominant SM processes that produce such events are \zjets and \ttbar,
which are simulated, respectively, using \ALPGEN \cite{Mangano:2002ea}
and \MADGRAPH~\cite{madgraph,Alwall:2007st} interfaced with \PYTHIA for
parton showering and hadronization.
Other backgrounds include multijet production
with two jets misidentified as electrons and $\PW$+jets events with one
jet mis-identified as an electron.
There is also a small contribution from di-boson and single top production.
The two leading (in \pt) electrons and two leading jets are used in the analysis and,
to reduce the backgrounds, required to have \pt $>$ 30 \pgev.
Selected electrons and jets have pseudorapidities
$|\eta|<2.5$ and $|\eta|<3.0$, respectively, and
if any of the selected electrons are closer than
$\Delta R = \sqrt{\Delta\eta^2 + \Delta\phi^2} = 0.7$
to any of the selected jets, the event is rejected.
In addition, the preliminary requirements $\mee>50$~GeV and \st$>$\ZbkgPrelimStCut are applied,
where \mee is the di-electron invariant mass
and \st is defined as the sum of the magnitudes of the \pt of the two leading electrons and two leading jets.
At this stage of the selection, referred to as pre-selection, there are sufficient data to
compare with the MC predictions.
Good data-MC agreement is observed in the shape of all kinematic
distributions of the selected electrons and jets.
Fig.~\ref{fig:MeeAfterPresel} shows the \mee and \st distributions.
The \zjets MC distributions have been normalized to the data
at the $\Z$ boson mass, as described later.

To reduce the background from \zee production, a minimal value of
\mee well above the mass of the $\Z$ boson is required, and,
to reduce all SM backgrounds, \st is required to be large.
While the LQ signal is expected to appear as a peak in the mass distribution of the electron-jet pairs,
we find that the $S_T$ variable is more powerful with the present
statistics as it is not affected by combinatorics.
The minimal values required for \mee and \st were optimized by minimizing
the expected upper limit on the leptoquark cross
section in the absence of an observed signal using a Bayesian approach
\cite{PDG2010,Bertram:2000br}
that is well suited for counting experiments in the Poisson regime.
The optimized lower value of \mee is found to be \MeeFinalCut~GeV
for all the LQ hypotheses under test, while the
lower value of \st varies as indicated in
Table~\ref{tab:EventSelSummary}.
Table~\ref{tab:EventSelSummary} shows the number of surviving events
for MC signal, MC background and data samples after
applying the full optimized selection.
The reported product of signal selection efficiency and acceptance is estimated from MC simulated events.
The product of the di-electron efficiency and acceptance,
prior to any \mee and jet requirements,
varies \twoEleEffic for LQ masses \LQMassesAnalyzed. 
\begin{table*}[htbp]
\caption{Number of events for MC LQ signal (for $\beta=1$), MC background,
and data samples after the full analysis selection and
corresponding to an integrated luminosity of \lumi.
The product of signal acceptance and efficiency is also reported for different LQ
masses.
The \zjets MC has been normalized to the data at the $\Z$ boson mass as described in the text.
Other backgrounds include $W$ + jets, di-boson, and single top.
Uncertainties are statistical. Systematic uncertainties are discussed later.
The observed and expected 95\% C.L. upper limit (u.l.) on the leptoquark
pair production cross section $\sigma$ are shown in the last column.}
\begin{center} \scriptsize 
\begin{tabular}{|c|cc|cccc|c|c|}
\hline\hline
$M_{\rm LQ}$    & \multicolumn{2}{|c|}{Signal Samples (MC)} & \multicolumn{4}{|c|}{Standard Model Background Samples (MC)}           & Events & Obs./Exp.\\
(\st~Cut)   & Selected        & Acceptance            & \multicolumn{4}{|c|}{Selected Events in}                       & in     & 95\% C.L.\\
{[GeV]}         & Events          & $\times$Efficiency    & $\ttbar$ + jets & $\Z/\gamma$ + jets & Others      & All      & Data   & u.l. on $\sigma$ [pb]\\
\hline
200 ($S_T>$340) & 117.5$\pm$0.8  & 0.297$\pm$0.002 & 2.2 $\pm$0.09 & 2.0 $\pm$0.2  & 0.27$\pm$0.05 & 4.5 $\pm$0.2  & 2 & 0.445 / 0.692 \\
250 ($S_T>$400) &  43.8$\pm$0.2  & 0.380$\pm$0.002 & 1.1 $\pm$0.06 & 1.3 $\pm$0.1  & 0.14$\pm$0.02 & 2.5 $\pm$0.1  & 1 & 0.311 / 0.442 \\
280 ($S_T>$450) &  24.4$\pm$0.1  & 0.403$\pm$0.002 & 0.59$\pm$0.05 & 0.87$\pm$0.07 & 0.10$\pm$0.02 & 1.6 $\pm$0.1  & 1 & 0.309 / 0.361 \\
300 ($S_T>$470) &  17.3$\pm$0.1  & 0.430$\pm$0.002 & 0.44$\pm$0.04 & 0.75$\pm$0.07 & 0.10$\pm$0.02 & 1.3 $\pm$0.1  & 1 & 0.294 / 0.326 \\
320 ($S_T>$490) &  12.3$\pm$0.1  & 0.451$\pm$0.002 & 0.37$\pm$0.04 & 0.65$\pm$0.07 & 0.08$\pm$0.02 & 1.1 $\pm$0.1  & 1 & 0.285 / 0.299 \\
340 ($S_T>$510) &  8.88$\pm$0.04 & 0.469$\pm$0.002 & 0.27$\pm$0.03 & 0.56$\pm$0.06 & 0.08$\pm$0.02 & 0.91$\pm$0.08 & 1 & 0.279 / 0.274 \\
370 ($S_T>$540) &  5.55$\pm$0.02 & 0.496$\pm$0.002 & 0.22$\pm$0.03 & 0.47$\pm$0.06 & 0.07$\pm$0.02 & 0.76$\pm$0.07 & 1 & 0.268 / 0.251 \\
400 ($S_T>$560) &  3.55$\pm$0.02 & 0.522$\pm$0.002 & 0.17$\pm$0.02 & 0.41$\pm$0.05 & 0.06$\pm$0.02 & 0.64$\pm$0.06 & 1 & 0.259 / 0.230 \\
450 ($S_T>$620) &  1.70$\pm$0.01 & 0.539$\pm$0.002 & 0.10$\pm$0.02 & 0.28$\pm$0.05 & 0.02$\pm$0.01 & 0.40$\pm$0.06 & 0 & 0.174 / 0.209 \\
500 ($S_T>$660) & 0.868$\pm$0.003& 0.565$\pm$0.002 & 0.07$\pm$0.02 & 0.23$\pm$0.05 & 0.02$\pm$0.01 & 0.32$\pm$0.05 & 0 & 0.166 / 0.193 \\
\hline\hline
\end{tabular}
\end{center}
\label{tab:EventSelSummary}
\end{table*}

The \zjets background dominates the pre-selection sample.
After the pre-selection, the ratio between data and MC events with $80<\mee<100$~GeV 
(where the contamination from other SM processes is \ContaminationAtZpeak) 
is \ZbkgRatioAtZPeak. This ratio is used to normalize the \zjets MC.
The statistical uncertainty on this normalization factor is used as an
uncertainty 
on the MC estimate of the \zjets background after the full selection.
The \ttbar background is estimated from MC with an uncertainty, 
\ttbarRelErr, taken from the uncertainty on the CMS measurement of the \ttbar cross 
section~\cite{Collaboration:2010ez}. Since this measurement is consistent
with NLO predictions, no rescaling of the \ttbar MC is applied.
The small contribution from other background processes containing vector bosons 
is estimated by MC.
The multijet background is determined from data. 
The probability that an isolated electromagnetic cluster is 
reconstructed as an electron is measured in a background sample requiring 
a single cluster, a jet multiplicity similar to 
the analysis final state, and small missing transverse energy.
This probability and a data sample with two or more of these clusters 
and two or more jets were used to determine the multijet contribution to the final 
selection sample.
The resulting systematic uncertainty is determined to be \uncertaintyQCD.
This background accounts for $<1\%$ of the total background for all LQ
masses hypotheses with a decreasing trend for increasing LQ mass hypothesis, 
and is not considered any further.

The systematic uncertainties affecting the number of
expected signal and background events are summarized
in Table~\ref{tab:SysUncertainties}.
The jet and electron energy scale uncertainties are given in the second column of Table~\ref{tab:SysUncertainties}.
The reconstruction, identification, and isolation efficiency for electrons
is determined from MC simulated events and a systematic uncertainty is assessed using
$Z\rightarrow ee$ events from collision data.
The statistical uncertainty on the number of MC events surviving
the full event selection is reported in Table~\ref{tab:EventSelSummary}
for the signal and background.
The uncertainty on the integrated luminosity of the data sample
is dominated by the uncertainty on the measurement of
the beam current~\cite{LumiNote}.
Uncertainties due to the choice of parton distribution functions (PDF)
of the proton lead to changes in the total cross section and the acceptance for both signal
and background processes.
The effect of the PDF uncertainties on the signal acceptance is estimated using an event re-weighting technique
that uses the LHAPDF package \cite{Bourilkov:2006cj} and amounts to 0.1\%.
Since a background normalization uncertainty is assessed based on data,
uncertainties due to the PDF choice, electron efficiencies, and integrated luminosity
are not applicable to the background estimate.

\begin{table*}[htbp]
\caption{Summary of the systematic uncertainties affecting the number of signal and background
events for the hypothesis of LQ with mass of \MLQforSystematics \GeVmass.}
\begin{center} \small
\begin{tabular}{|l|c|c|c|}
\hline\hline
Systematic  & Magnitude & Effect on            & Effect on            \\
Uncertainty & $[\%]$    & Signal [\%]         & Background [\%]  \\
\hline
Background Normalization Uncertainty &-& - & 21 \\
Jet Energy Scale &5& 3& 11\\
Electron Energy Scale Barrel/Endcap &1/3& 1& 5\\
Electron Pair Reco, ID, Iso &10& 10& - \\
MC Statistics & - & 1& 6\\
Integrated Luminosity &11& 11& - \\
\hline
Total & - &15 & 25\\
\hline\hline
\end{tabular}
\label{tab:SysUncertainties}
\end{center}
\end{table*}

\begin{figure}[htbp]
  \centering
  \includegraphics[width=0.4\textwidth]{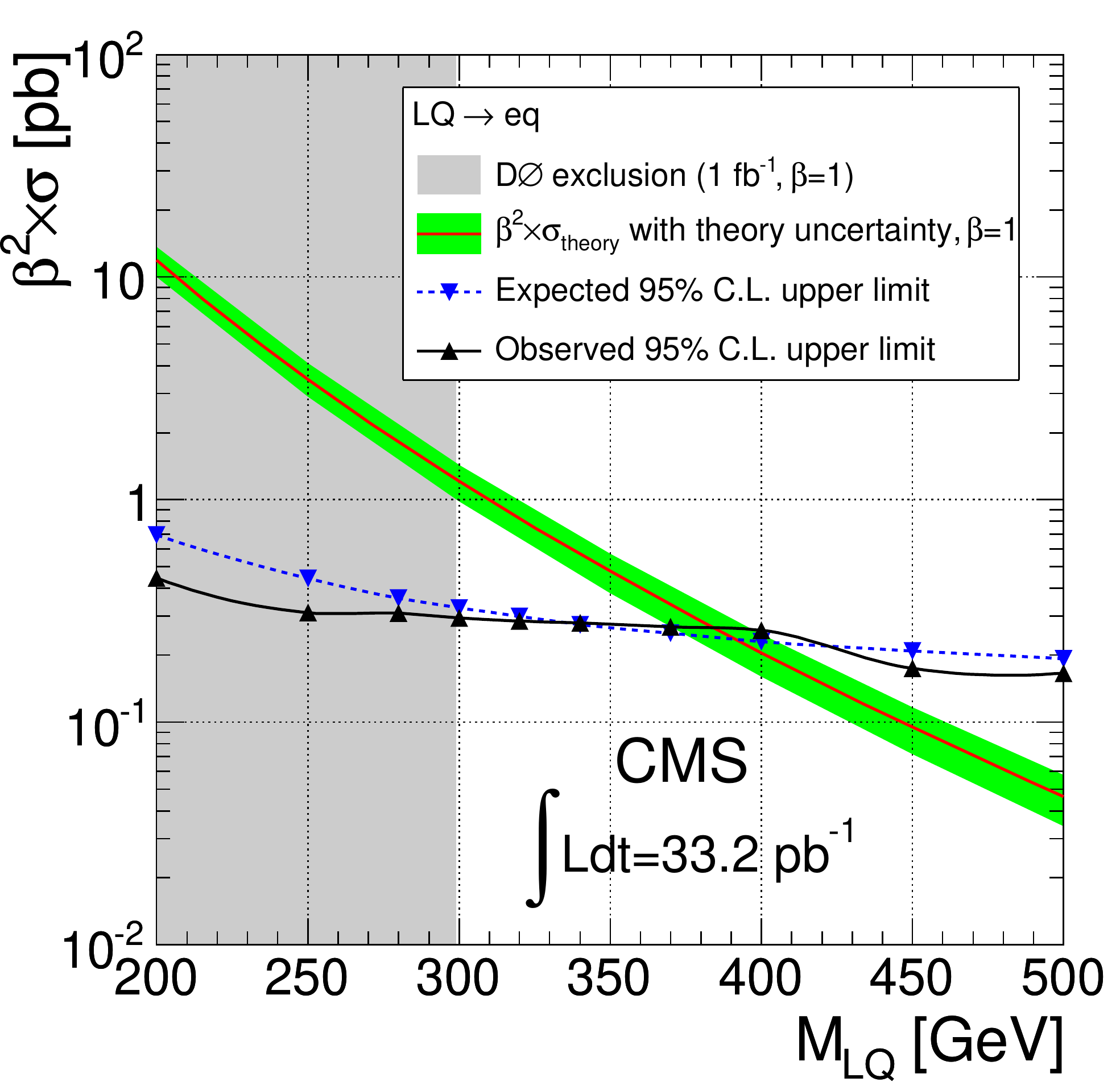}
  \includegraphics[width=0.4\textwidth]{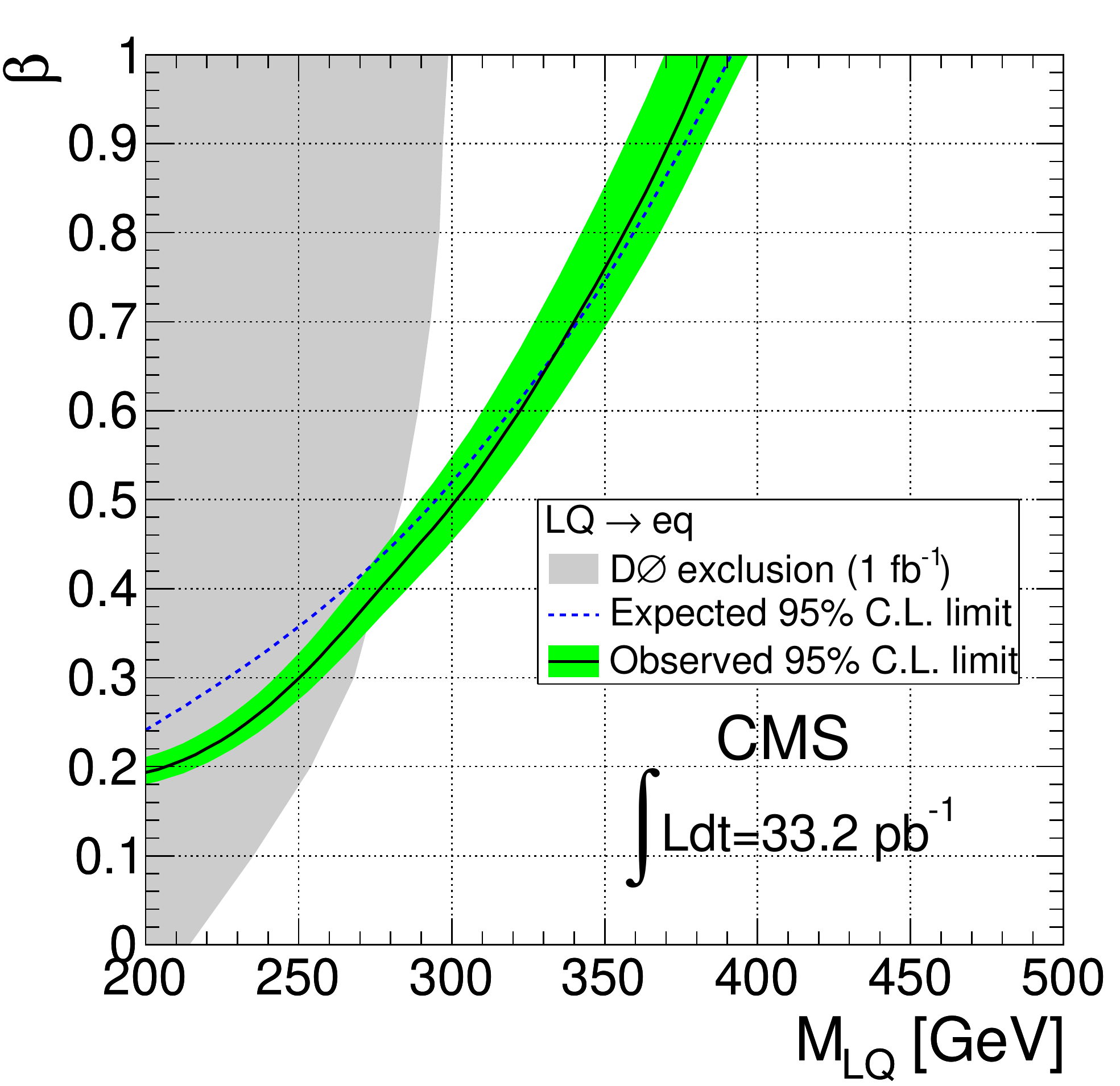}
  \caption{
    On the left:
    the expected and observed upper limit at $95\%$ C.L. on the LQ pair production
    cross section times $\beta^2$ as a function of the LQ mass.
    The systematic uncertainties reported in Table~\ref{tab:SysUncertainties} are included in the calculation.
    The shaded region is excluded by the current \DZERO limit for $\beta=1$.
    The $\sigma_\text{theory}$ curve and its band represent, respectively, the theoretical LQ pair production cross section
    and the uncertainties due to the choice of PDF and renormalization/factorization scales \cite{PhysRevD.71.057503}.
    On the right:
    minimum $\beta$ for a 95\% C.L. exclusion of the LQ hypothesis as a function of LQ mass.
    The observed (expected) exclusion curve is obtained using the
    observed (expected) upper limit and the central value of the theoretical LQ pair production cross section.
    The band around the observed exclusion curve is obtained by considering the observed upper limit
    while taking into account the uncertainties on the theoretical cross section.
    The shaded region is excluded by the current \DZERO limits, which combines results from searches in the two electron, electron-neutrino,
and two neutrino channels.
    }
  \label{fig:exclusion_xs}
\end{figure}

The number of observed events in the collision data sample
that pass the selection criteria optimized for each LQ mass considered
is consistent with the prediction from SM processes,
as reported in Table~\ref{tab:EventSelSummary}.
An upper limit on the LQ cross section in the absence of signal
is therefore set using a Bayesian approach
\cite{Bertram:2000br}
that uses a Poisson likelihood, a flat prior for
the signal cross section, and log-normal priors for the
parameters used to model the systematic uncertainties.
Systematic uncertainties for the signal are dominated
by the uncertainty on the integrated luminosity and the electron selection efficiencies,
while the systematic uncertainties for the background are dominated by
the uncertainty derived from data. 
Fig.~\ref{fig:exclusion_xs} (left)
shows the 95\% Confidence Level (C.L.)
upper limit on the LQ pair production cross section times $\beta^2$ as a
function of the leptoquark mass for \lumi of integrated luminosity.
The systematic uncertainties reported
in Table~\ref{tab:SysUncertainties} are included in the calculation.
The upper limits are compared to an NLO prediction
of the LQ pair production cross section \cite{PhysRevD.71.057503} to set a 95\% C.L. exclusion
on LQ masses smaller than \LQmassLimitBetaOne (expected 391~GeV), assuming $\beta=1$.
A theoretical uncertainty on the signal production cross sections
due to the choice of renormalization/factorization scales
(\theoreticalUncertaintyOnLQCrossSection)
has been calculated by varying the scales between half and twice the LQ mass,
while a 90\% C.L. PDF uncertainty (\PDFUncertaintyOnLQCrossSection)
has been obtained from the CTEQ6.6 
error PDF set following the standard prescription detailed in Ref.~\cite{Nadolsky:2008zw}.
If the observed cross section upper limit is compared with the lower boundary of the cross section uncertainty band,
the lower limit on the LQ mass for $\beta=1$ becomes \LQmassLimitBetaOneWithTheoryError (expected $375$~GeV).
Fig.~\ref{fig:exclusion_xs} (right) shows the minimum $\beta$ for a 95\% C.L. exclusion
of the LQ hypothesis as a function of LQ mass.

In conclusion, a search for pair production of first-generation scalar leptoquarks has been presented.
The number of collision events, passing a selection optimized for 
exclusion of the LQ hypothesis, is in good agreement with the predictions 
for the SM background processes.
A Bayesian approach 
that includes the treatment of the systematic uncertainties as nuisance parameters 
has been used to set an upper limit on the LQ cross section. 
By comparing this upper limit to a theoretical calculation of the LQ pair production 
cross section, the existence of first-generation scalar LQ with masses below 
\LQmassLimitBetaOne for $\beta=1$ has been excluded at 95\% C.L.,
with a corresponding cross section limit of 0.267 pb. 
The lower limits on the LQ mass set for values of $\beta$ larger than about 0.4 
are the most restrictive direct limits to date.
We wish to thank Michael Kr\"amer for providing the NLO LQ pair production 
cross sections at $\sqrt{s}=7$~TeV. 
We congratulate our colleagues in the CERN accelerator
departments for the excellent performance of the LHC machine. We thank
the technical and administrative staff at CERN and other CMS
institutes, and acknowledge support from: FMSR (Austria); FNRS and FWO
(Belgium); CNPq, CAPES, FAPERJ, and FAPESP (Brazil); MES (Bulgaria);
CERN; CAS, MoST, and NSFC (China); COLCIENCIAS (Colombia); MSES
(Croatia); RPF (Cyprus); Academy of Sciences and NICPB (Estonia);
Academy of Finland, ME, and HIP (Finland); CEA and CNRS/IN2P3
(France); BMBF, DFG, and HGF (Germany); GSRT (Greece); OTKA and NKTH
(Hungary); DAE and DST (India); IPM (Iran); SFI (Ireland); INFN
(Italy); NRF and WCU (Korea); LAS (Lithuania); CINVESTAV, CONACYT,
SEP, and UASLP-FAI (Mexico); PAEC (Pakistan); SCSR (Poland); FCT
(Portugal); JINR (Armenia, Belarus, Georgia, Ukraine, Uzbekistan); MST
and MAE (Russia); MSTD (Serbia); MICINN and CPAN (Spain); Swiss
Funding Agencies (Switzerland); NSC (Taipei); TUBITAK and TAEK
(Turkey); STFC (United Kingdom); DOE and NSF (USA).

\bibliography{auto_generated}   
\cleardoublepage\appendix\section{The CMS Collaboration \label{app:collab}}\begin{sloppypar}\hyphenpenalty=5000\widowpenalty=500\clubpenalty=5000\textbf{Yerevan Physics Institute,  Yerevan,  Armenia}\\*[0pt]
V.~Khachatryan, A.M.~Sirunyan, A.~Tumasyan
\vskip\cmsinstskip
\textbf{Institut f\"{u}r Hochenergiephysik der OeAW,  Wien,  Austria}\\*[0pt]
W.~Adam, T.~Bergauer, M.~Dragicevic, J.~Er\"{o}, C.~Fabjan, M.~Friedl, R.~Fr\"{u}hwirth, V.M.~Ghete, J.~Hammer\cmsAuthorMark{1}, S.~H\"{a}nsel, C.~Hartl, M.~Hoch, N.~H\"{o}rmann, J.~Hrubec, M.~Jeitler, G.~Kasieczka, W.~Kiesenhofer, M.~Krammer, D.~Liko, I.~Mikulec, M.~Pernicka, H.~Rohringer, R.~Sch\"{o}fbeck, J.~Strauss, A.~Taurok, F.~Teischinger, W.~Waltenberger, G.~Walzel, E.~Widl, C.-E.~Wulz
% ...
\vskip\cmsinstskip
% ...
1:~~Also at CERN, European Organization for Nuclear Research, Geneva, Switzerland\\
% ...

\end{sloppypar}
\end{document}